\documentclass{sig-alternate-05-2015}
\usepackage{url}

\usepackage{multirow}
\usepackage{array}

\newcolumntype{L}[1]{>{\raggedright\let\newline\\\arraybackslash\hspace{0pt}}m{#1}}
\newcolumntype{C}[1]{>{\centering\let\newline\\\arraybackslash\hspace{0pt}}m{#1}}
\newcolumntype{R}[1]{>{\raggedleft\let\newline\\\arraybackslash\hspace{0pt}}m{#1}}

\usepackage{balance}
\usepackage{rotating}
\usepackage{booktabs}
\usepackage{hyphsubst}
\usepackage{graphicx}
\usepackage{amsmath}
\graphicspath{{images/}}
\usepackage{times}
\usepackage{hhline}
\usepackage[caption = false]{subfig}

\usepackage{url}
\usepackage{rotating}
\usepackage{amsmath}
\usepackage{cancel}
\usepackage{amssymb}%
\usepackage{enumitem}
\usepackage{pifont}%

\usepackage{multirow}
\usepackage{array}
\usepackage{placeins}

\usepackage[dvipsnames]{xcolor}
\usepackage{listings}

\usepackage{tabularx}

\usepackage{booktabs}
\usepackage{siunitx}
\newcolumntype{L}{@{}l@{}} %
\newcommand{\mc}[1]{\multicolumn{1}{c}{#1}} %

\newcommand\YAMLcolonstyle{\color{orange}\mdseries}
\newcommand\YAMLkeystyle{\color{black}\bfseries}
\newcommand\YAMLvaluestyle{\color{blue}\mdseries}

\makeatletter

\newcommand\language@yaml{yaml}

\expandafter\expandafter\expandafter\lstdefinelanguage
\expandafter{\language@yaml}
{
  keywords={true,false,null,y,n},
  keywordstyle=\color{blue},
  basicstyle=\YAMLkeystyle,                                 %
  sensitive=false,
  comment=[l]{\#},
  morecomment=[s]{/*}{*/},
  commentstyle=\color{red}\ttfamily,
  stringstyle=\YAMLvaluestyle\ttfamily,
  moredelim=[l][\color{orange}]{\&},
  moredelim=[l][\color{magenta}]{*},
  moredelim=**[il][\YAMLcolonstyle{:}\YAMLvaluestyle]{:},   %
  morestring=[b]',
  morestring=[b]",
  literate =    {---}{{\ProcessThreeDashes}}3
                {>}{{\textcolor{orange}\textgreater}}1     
                {|}{{\textcolor{orange}\textbar}}1 
                {\ -\ }{{\mdseries\ -\ }}3,
}

\lst@AddToHook{EveryLine}{\ifx\lst@language\language@yaml\YAMLkeystyle\fi}
\makeatother

\newcommand\ProcessThreeDashes{\llap{\color{cyan}\mdseries-{-}-}}

\definecolor{pblue}{rgb}{0.13,0.13,1}
\definecolor{pgreen}{rgb}{0,0.5,0}
\definecolor{pred}{rgb}{0.9,0,0}
\definecolor{pgrey}{rgb}{0.46,0.45,0.48}

\usepackage{rotating}
\usepackage{booktabs}
\usepackage{rotating}
\usepackage{hhline}
\usepackage{hyphsubst}
\usepackage{soul}

\newlength{\Oldarrayrulewidth}

\begin{document}

\title{High Enough? Explaining and Predicting Traveler Satisfaction Using Airline Reviews}

\numberofauthors{3}
\author{
\alignauthor
Emanuel Lacic\\
       \affaddr{KTI}\\
       \affaddr{Graz University of Technology}\\
       \affaddr{Graz, Austria}\\
       \email{elacic@know-center.at}
\alignauthor
Dominik Kowald\\
       \affaddr{Know-Center}\\
       \affaddr{Graz University of Technology}\\
       \affaddr{Graz, Austria}\\
       \email{dkowald@know-center.at}
\alignauthor
Elisabeth Lex\\
        \affaddr{KTI}\\
       \affaddr{Graz University of Technology}\\
       \affaddr{Graz, Austria}\\
       \email{elisabeth.lex@tugraz.at}
}

\permission{}
\conferenceinfo{HT'16,}{July 10--13, 2015, Halifax, Canada.}
\copyrightetc{\copyright 2016 ACM. \the\acmcopyr}
\crdata{ISBN xxx-x-xxxx-xxxx-x/xx/xx...\$xx.xx \\
DOI: http://dx.doi.org/xx.xxxx/xxxxxxx.xxxxxxx}

\ccsdesc[500]{Information systems~Clustering and classification}

\maketitle

\begin{abstract}
Air travel is one of the most frequently used means of transportation in our every-day life. Thus, it is not surprising that an increasing number of travelers share their experiences with airlines and airports in form of online reviews on the Web. In this work, we thrive to explain and uncover the features of airline reviews that contribute most to traveler satisfaction. To that end, we examine reviews crawled from the Skytrax air travel review portal. Skytrax provides four review categories to review airports, lounges, airlines and seats. Each review category consists of several five-star ratings as well as free-text review content. In this paper, we conducted a comprehensive feature study and we find that not only five-star rating information such as airport queuing time and lounge comfort highly correlate with traveler satisfaction but also textual features in the form of the inferred review text sentiment. Based on our findings, we created classifiers to predict traveler satisfaction using the best performing rating features. Our results reveal that given our methodology, traveler satisfaction can be predicted with high accuracy. Additionally, we find that training a model on the sentiment of the review text provides a competitive alternative when no five star rating information is available. We believe that our work is of interest for researchers in the area of modeling and predicting user satisfaction based on available review data on the Web.

\end{abstract}

\printccsdesc

\keywords{traveler satisfaction; airline reviews; skytrax; user satisfaction prediction; feature analysis; sentiment analysis; clustering analysis}

\section{Introduction}
\label{sec:introduction}
In the last decades, air travel has become one of the most frequently used means of transportation. The International Air Transport Association (IATA) expects traveler numbers to reach 7.3 billion by 2034, representing a 4.1\% average annual growth in demand for air connectivity\footnote{http://www.iata.org/pressroom/pr/Pages/2014-10-16-01.aspx}. At the same time, an increasing number of airlines is competing for market shares, which raises the need to attract customers while balancing costs and services.

An increasing number of customers (i.e., travelers) share their experiences and viewpoints on airlines and airports in form of online reviews in order to help others to better judge airline and airport quality. Such reviews may consist of free-text reviews combined with ratings (e.g., by means of 5-star ratings). 
As a consequence, a vast amount of airline review data is available on the Web, which is not only of interest for the airline industry but also for researchers working on analyzing the impact of the factors/features contributing to user satisfaction \cite{chatterjee2001online,chen2004impact,chen2003marketing}. 

In this paper, we present work-in-progress on a recently started project that aims at explaining and predicting traveler satisfaction using airline review data. Specifically, it is our goal to identify critical features that contribute to air travel satisfaction based on rating and textual reviews. Our idea is to exploit these features in order to predict whether a traveler is satisfied with her airline/airport choice based on the given ratings and/or textual review. This is summed up in the following two research questions that guide our work:
\begin{itemize}
\item[\textit{RQ1:}] Which rating and textual features of airline reviews are most indicative for traveler satisfaction?
\item[\textit{RQ2:}] To what extent can we predict traveler satisfaction using the available rating and textual features of airline reviews?
\end{itemize}

\begin{figure}[t!]
        \centering
  		 \subfloat[Review text]{ 
				\includegraphics[width=0.22\textwidth]{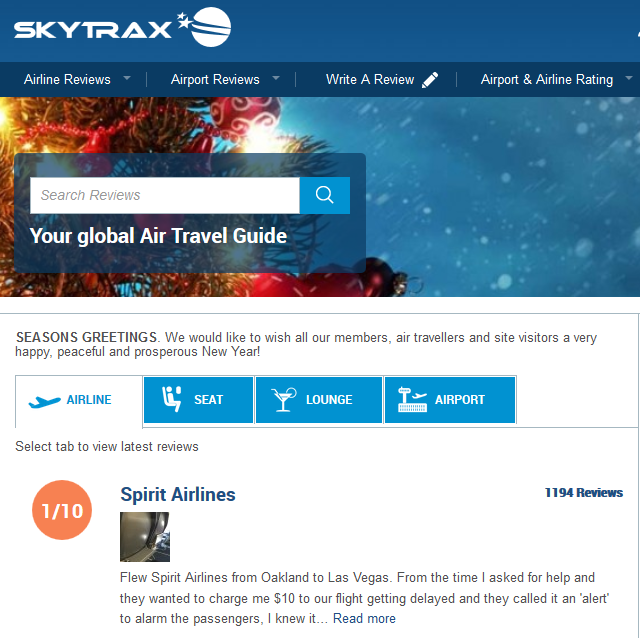}%
		 }
		 \subfloat[Ratings]{ 
				\includegraphics[width=0.22\textwidth]{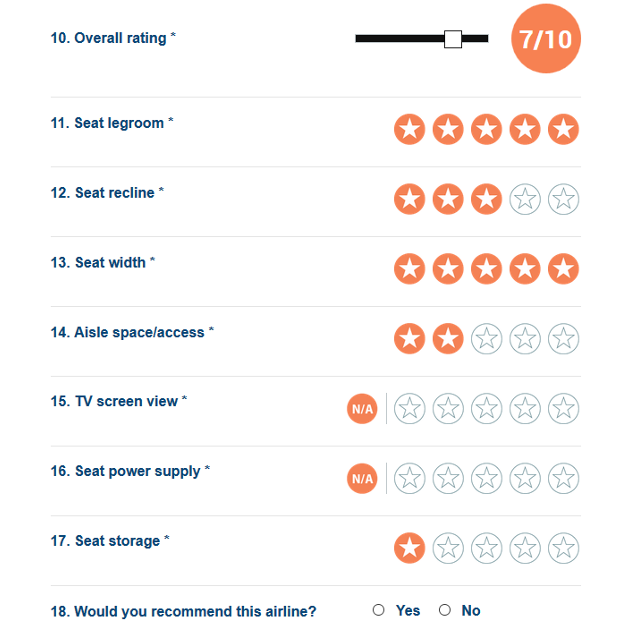}%
		 }
     \caption{The Skytrax airline (a) review and (b) rating portal. Within each of the four review categories, users state their traveler satisfaction via several rating features, a review text, an overall rating and a binary signal indicated by the \textit{Would you recommend this airline/airport?} checkbox.\vspace{-3mm}}
    \label{fig:skytrax}
\end{figure}

\noindent \textbf{Explaining traveler satisfaction.}
In order to better explain how the features contribute to traveler satisfaction, and thus, to address \textit{RQ1}, we exploit real-world airline review data, which was crawled from the Skytrax portal. As shown in Figure \ref{fig:skytrax}, in Skytrax users can (a) enter review text and (b) rate various services. Moreover, the user can state her final traveler satisfaction not only using an overall rating between 1 and 10 but also using a checkbox to indicate if she would recommend this airline or airport to other travelers. In terms of rating features, we explore features derived from four different review categories, namely airport, lounge, airline and seat reviews. In terms of textual features, we infer the sentiment of the review text (see Section \ref{sec:dataset}).

To identify which review features are most indicative for traveler satisfaction, we conduct a feature analysis in which we correlate rating and textual features with the overall rating given by the user. We find that airport \textit{queuing} time, lounge \textit{comfort}, airline \textit{cabin staff} quality and seat \textit{legroom} space are factors that highly impact the overall traveler satisfaction. We also find that the sentiment of the review content is a good indicator to determine whether a traveler was satisfied with the travel. Additionally, we perform clustering and cluster labelling of the textual content in order to identify topics which are discussed in the reviews. In the long run, this may help to extend the rating schema. For example, if many users discuss the topic ``immigration'' in their textual review, the rating portal could introduce a novel rating feature, which enables users to rate the quality of the immigration service.

\vspace{2mm}
\noindent \textbf{Predicting traveler satisfaction.}
We utilize the available rating information as well as the sentiment of the textual reviews as features for our prediction study (\textit{RQ2}). We formulate the prediction task as a binary classification problem of the final traveler satisfaction signal indicated by the \textit{Would you recommend this airline/airport?} checkbox (see Figure \ref{fig:skytrax}).

We find strong performance in predicting the traveler satisfaction using the individual rating features. By using a combination of the best performing rating features, we demonstrate that the prediction accuracy can even be increased. Additionally, we show that a classifier, which solely uses the sentiment of the review text, provides a competitive performance in terms of prediction accuracy as well. This could especially be beneficial in cases where rating features are missing. In terms of metrics, we report the prediction accuracy by means of the F1-score and AUC (i.e., area under ROC curve).

\vspace{2mm}
\noindent \textbf{Significance of this work.} 
With this study, we aim at explaining which rating and textual features of airline reviews have the most impact on predicting traveler satisfaction. Our findings can provide guidance for stakeholders in the airline industry, as well as for researchers, who study online review data to better understand what is important to travellers and what impacts user satisfaction.

\section{Related work}
\label{sec:related_work}
Since Heskett et al. \cite{heskett1994putting} established a relationship between traveler satisfaction and profitability, research on the airline service quality has become an important issue for the airline industry. 
As a consequence, the authors of \cite{saha2009service} claim that it is crucial to continuously collect and evaluate data about traveler satisfaction and how it relates to the provided service quality in order to be competitive in the airline industry.
However, most work that conduct research in airline service quality rely on gathered offline data coming from on-site questionnaires \cite{suki2014passenger, liou2010dominance}, airline submissions \cite{tiernan2008airline} or in-depth interviews \cite{Vlachos2014}. 

Nowadays, online reviews are getting more popular and as a consequence there is the opportunity to leverage them as a rich and powerful source of information. In fact, there is a lot of valuable hidden information available in online reviews \cite{pang2008opinion}. As such, Web sites like the already mentioned Skytrax portal, AirlineRatings\footnote{\url{http://http://www.airlineratings.com/}} and TripAdvisor\footnote{\url{http://www.tripadvisor.com}} are important for the airline industry to explain how service quality is perceived by the travelers. Furthermore, this data may be a valuable source for researchers that aim at better understanding the factors that contribute to user satisfaction.

One recent work going into that direction is the one described in \cite{Yakut2015}, in which the authors mined review data about airlines' in-flight services from the Skytrax portal. By grouping travelers via feature-based and clustering-based modelling, the authors showed that inferences can be captured to explain how travelers evaluate in-flight services. 
Another recent work of Yao et al. \cite{yao2015exploring} presented a research framework to extract and explore information on a user's opinion about airline service features from a large static corpus of online review texts. 

In our work, we perform a comprehensive feature analysis using rating and textual features from airport, lounge, airline and seat reviews in order to explain which features actually contribute to traveler satisfaction. Moreover, we show how the different rating and textual features can be utilized to predict traveler satisfaction. Our methods and results provide practical insights on how to build upon work like \cite{yao2015exploring} in order to predict traveler satisfaction using online airline reviews.

\begin{table}[t!]
\setlength\extrarowheight{1.4pt}
\centering
\small
\begin{tabular}{l|c|c|c|c}
\hline \hline
Review categories & Airports & Lounges & Airlines & Seats \\ \hline \hline
\# Users                   & 11,834  & 1,598         & 29,645 & 1,147      \\ 
\# Reviews                 & 17,721  & 2,264         & 41,396 & 1,258      \\ \hline
Traveler Satisfaction      & 22.12\%  & 36.04\%    & 53.38\% & 36.41\%  \\ \hline \hline
\end{tabular}
\caption{Statistics of the Skytrax dataset showing how many reviews were given by the users in the four categories. Additionally, we report the traveler satisfaction in the categories as the relative number of reviews that were indicated as airlines or airports that would be recommended to other travelers.\vspace{-4mm}}
\label{tab:data}
\end{table}

\begin{figure*}[t!]
        \centering
  		 \subfloat[Airport reviews]{ 
				\includegraphics[width=0.25\textwidth]{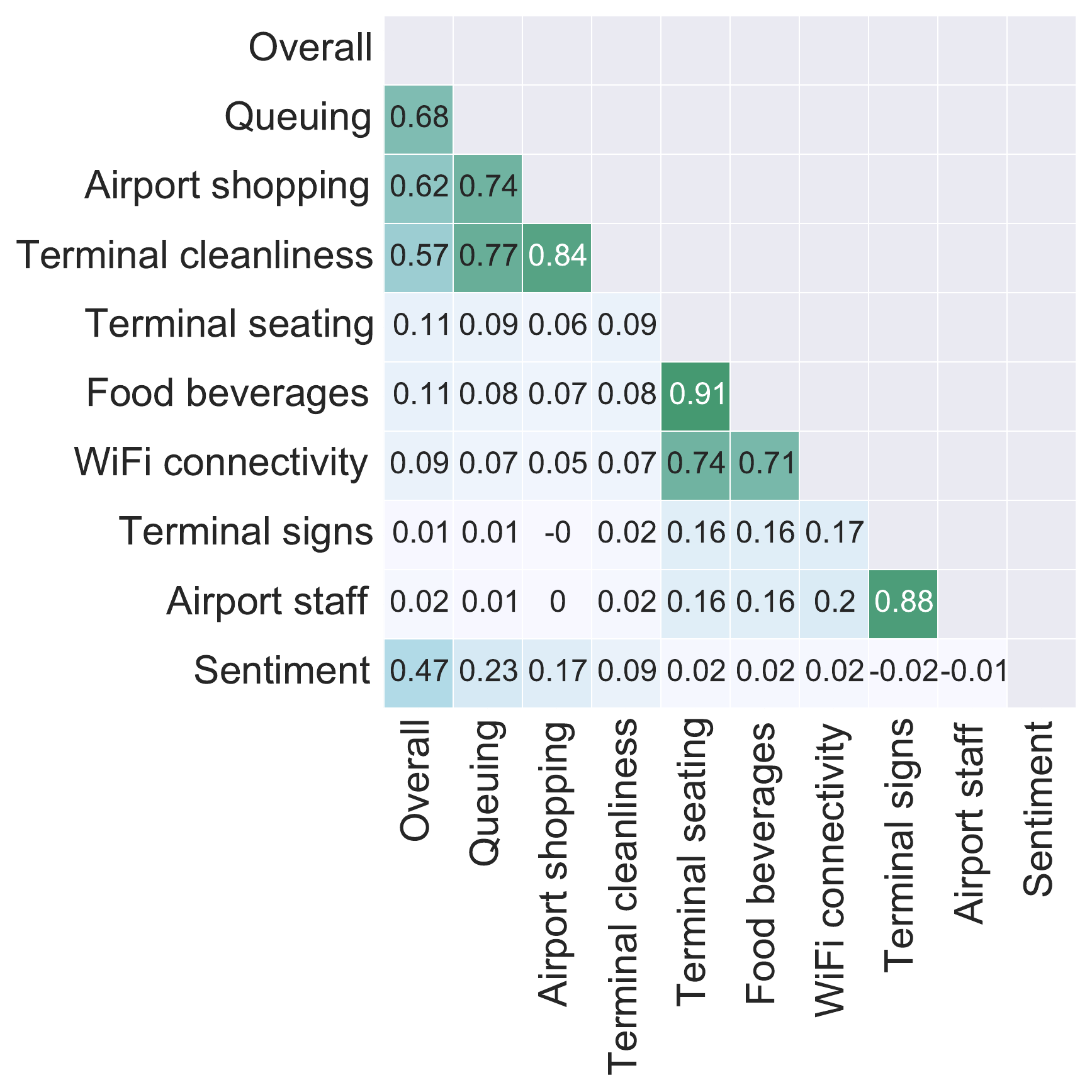}%
		 }
		 \subfloat[Lounge reviews]{ 
				\includegraphics[width=0.25\textwidth]{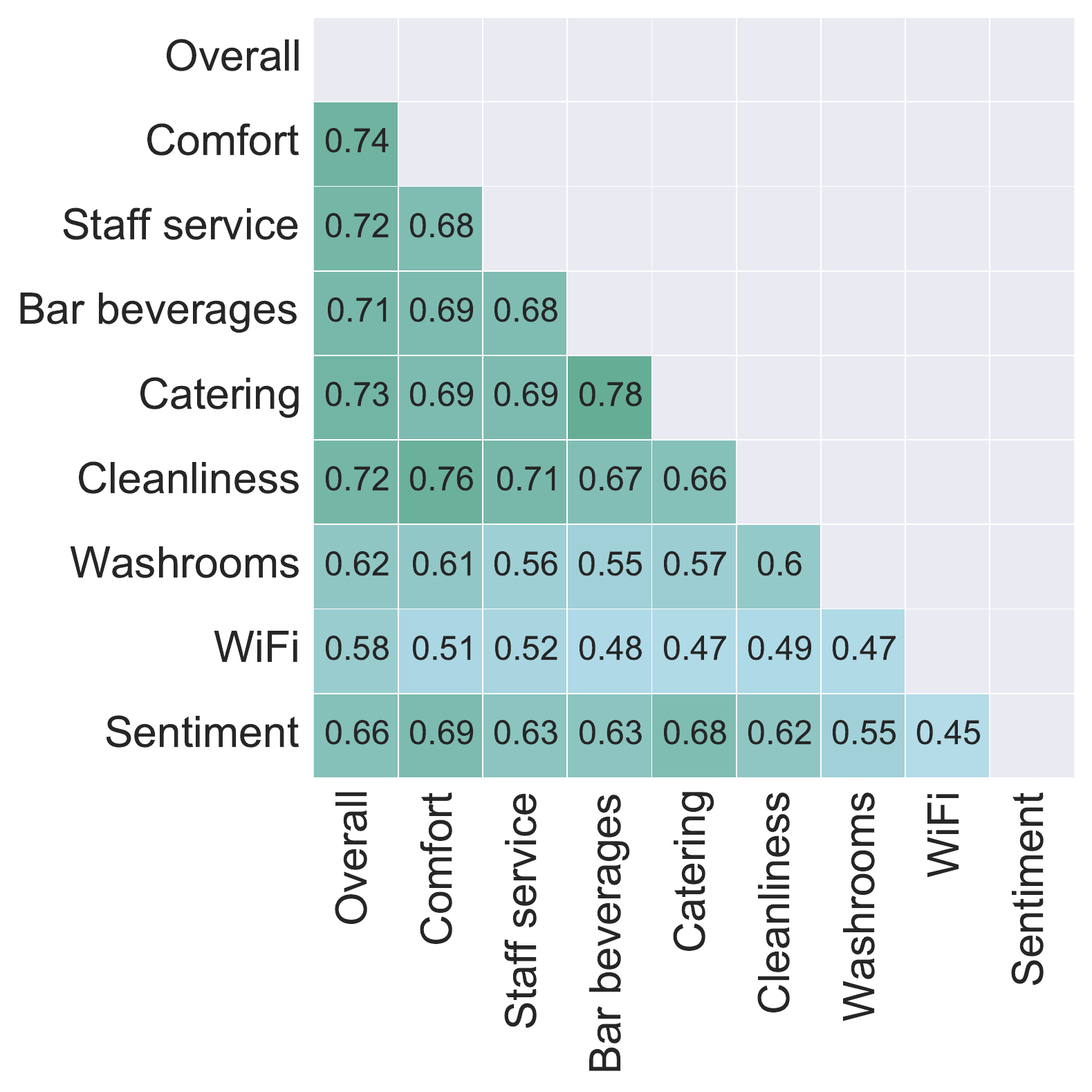}%
		 }
	 \subfloat[Airline reviews]{ 
				\includegraphics[width=0.25\textwidth]{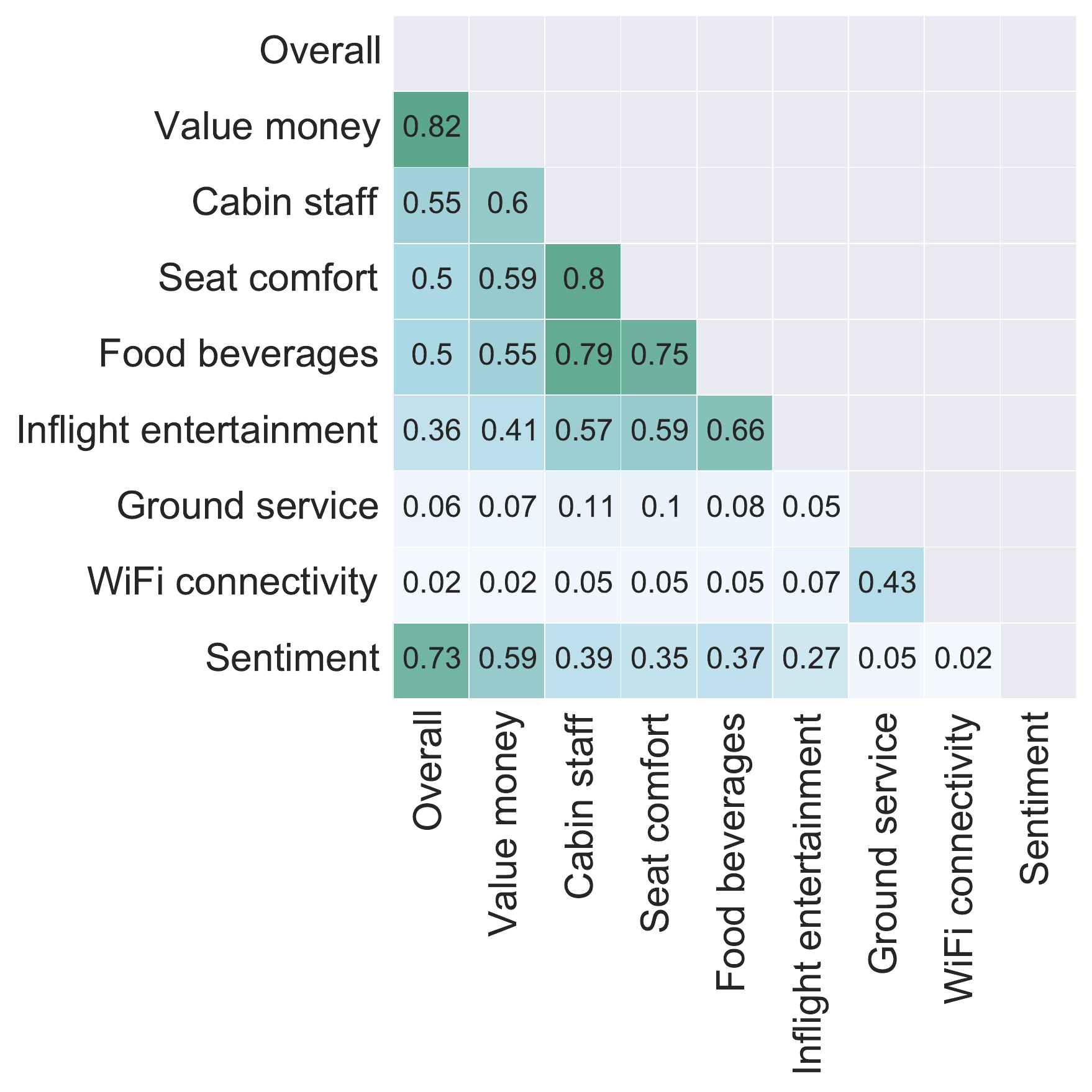}%
		 }
		 	 \subfloat[Seat reviews]{ 
				\includegraphics[width=0.25\textwidth]{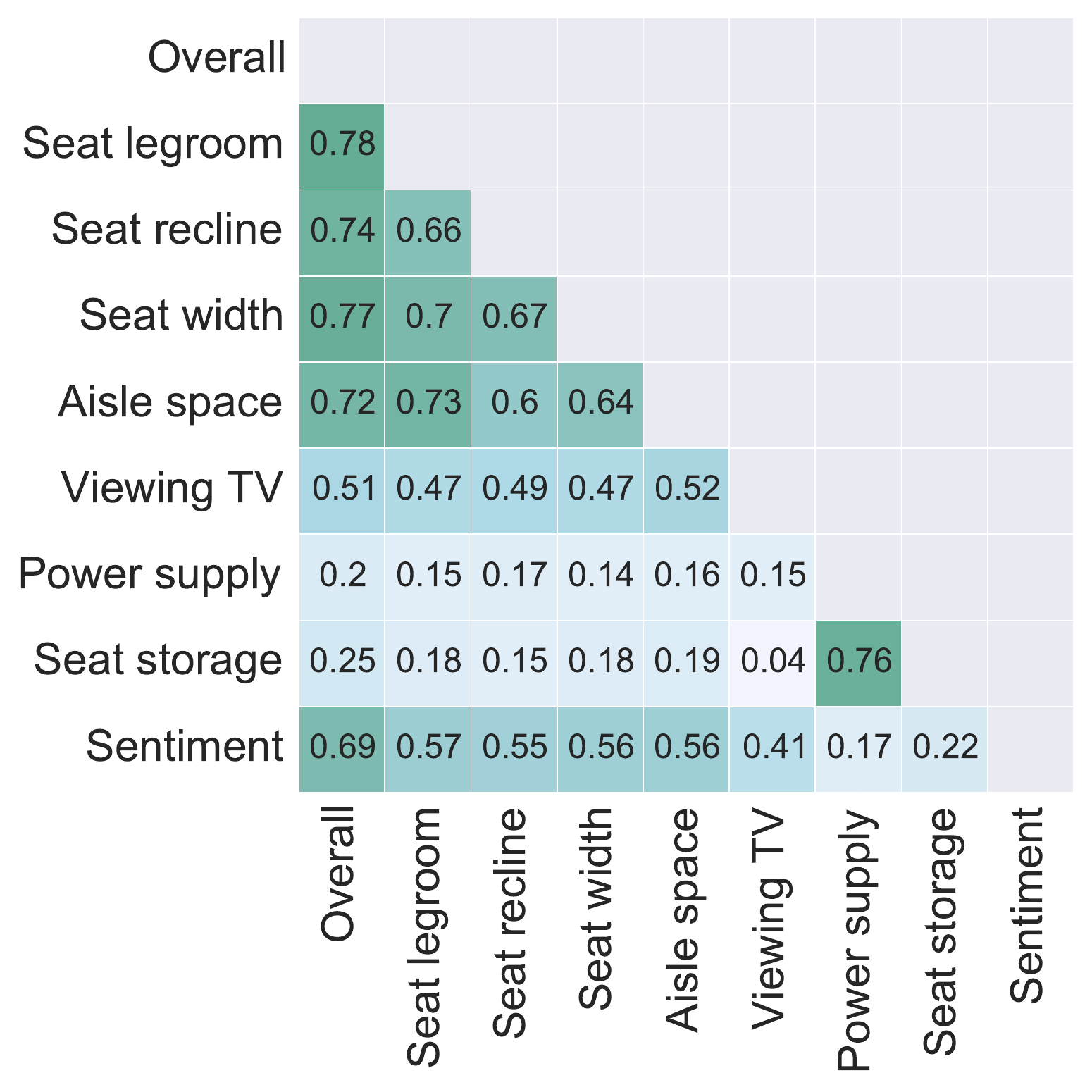}%
		 }
     \caption{Pearson correlation of the rating and textual features with the overall rating given by the users. Besides the rating and textual features (i.e., sentiment), each review category shows the overall rating indicating traveler satisfaction (\textit{RQ1}). \textit{Note:} all correlations values higher than .02 have a p-value < .000001.\vspace{-4mm}}
    \label{fig:correlation}
\end{figure*}

\section{Airline Review Data}
\label{sec:dataset}
Within the air travel industry, the London-based company Skytrax has established itself as a leader in conducting air travel research. 
Skytrax provides international audits and airport rankings and gives traveler-based satisfaction awards in its yearly \textit{World Airport Awards} and \textit{World Airline Awards}. 
Their airport and airline review Web portal has positioned itself as one of the most popular independent review sites within the air travel industry.
In this work, we incorporate a recent publicly made available airline review dataset\footnote{\url{https://github.com/quankiquanki/skytrax-reviews-dataset}} scraped from Skytrax's Web portal. This dataset contains not only rating and textual features of airline reviews but also features that indicate the final traveler satisfaction (see Figure \ref{fig:skytrax}).

\vspace{2mm} \noindent \textbf{Rating features.} 
The rating data gathered from Skytrax is divided into four different review categories: (1) airport, (2) lounge, (3) airline, and (4) seat reviews. Each review category has 7 - 8 individual rating features that map the perceived quality of a specific service. The individual rating features are based on a 5-star scale and are accompanied by an additional overall rating on a 1 - 10 scale. Table \ref{tab:data} shows the statistics of the dataset and reveals that most reviews are targeted at airports and airlines, and less at specific seats or lounges. 

\vspace{2mm} \noindent \textbf{Textual features.} 
The posted review text can also contain valuable information about the perceived service quality and satisfaction of a traveler \cite{pang2008opinion}. To that end, we manually enriched the available dataset by inferring the sentiment of each review text. Based on recent research, which compared several sentiment analysis tools \cite{serrano2015sentiment}, we extracted the additional textual feature using AlchemyAPI\footnote{\url{http://www.alchemyapi.com/}}. As we will show in this paper, the sentiment of the review text further helps in explaining and predicting traveler satisfaction and is especially useful when rating features are missing.

\vspace{2mm} \noindent \textbf{Traveler Satisfaction.}
We use the overall rating to evaluate how the different rating and textual features influence the traveler's satisfaction. Furthermore, in order to make a final decision on how a traveler was satisfied with an airline or airport, we utilize the binary signal represented as the \textit{Would you recommend this airline/airport?} checkbox of Skytrax. As such, Table \ref{tab:data} also shows how travelers are satisfied based on the four review categories. For example, airport reviews mostly resulted in a negative traveler satisfaction, whereas airline reviews almost contain the same amount of satisfactory and unsatisfactory experiences. %

\section{Explaining \\ Traveler Satisfaction}
\label{sec:fe}
In this section, we aim to answer the first research question of our work (\textit{RQ1}) and determine the rating and textual features that contribute the most to traveler satisfaction.

\subsection{Methodology}
As already outlined in Section \ref{sec:dataset}, each review category reveals an overall rating, which states how a user perceived an airport, lounge, airline or seat during the travel. 
For example, the \textit{Dalaman} airport, located in south-west Turkey, received the worst overall rating with a mean of $2.17$. On the contrary, the best rated airport is the \textit{Singapore's Changi} airport with an average overall rating of $7.09$.
With respect to airlines, \textit{Bangkok Airways} was the best rated one with a mean overall rating of $7.99$, whereas \textit{Air Canada rouge} is the worst rated airline with a mean of $2.54$. 

In order to determine which features actually influence these overall scores, we conduct a feature analysis in which we correlate the rating and textual features (i.e., the sentiment) with the overall rating given by the user. To explore the influences of rating and textual features, we use the Pearson's product-moment correlation coefficient \cite{lee1988thirteen}. In this respect, we further correlate the ratings of the features among each other because we believe that knowing how features influence not only the overall rating but also the rating of other features, helps us in even better understanding the factors that contribute to traveler satisfaction.

In addition to the correlation analysis of rating and textual features, we further incorporate the textual content of online airline reviews. Our aim is to uncover additional features that could be introduced to the rating schema. To that end, we perform clustering and cluster labeling of the review content in order to identify topics which are discussed in reviews. In contrast to \cite{Yakut2015}, we do not cluster the content with the commonly used k-means approach but rather using Suffix Tree Clustering (STC) \cite{Zamir1998}, an approach that focuses on the problem of cluster labeling. We justify our choice since this clustering technique merges base clusters with high textual overlaps and was shown to outperform group average agglomerative hierarchical clustering, k-means, buckshot, fractionation and single-pass algorithms \cite{Zamir1998, sambasivam2006advanced}.

\subsection{Results}
Figure \ref{fig:correlation} shows the results of our feature correlation analysis on rating and textual (i.e., sentiment) features based on the four categories.

\vspace{2mm} \noindent \textbf{Airport reviews.}
In airport reviews, the \textit{overall rating} is mostly influenced by (long) \textit{queuing} times, quality of \textit{airport shopping} and the \textit{cleanliness of the terminal}. A mild correlation with the sentiment of the review text can also be found. One interesting observation is that traveler satisfaction about \textit{terminal seats} is heavily influenced by the offered \textit{foods and beverages} as well as available \textit{WIFI connectivity}. It can also be observed that travelers will not be pleased with the \textit{airport staff} when they are experiencing issues with the \textit{terminal signs}. 

\vspace{2mm} \noindent \textbf{Lounge reviews.} Compared to airport reviews, the overall satisfaction within lounge reviews is highly influenced by most rating features. The top four indicators are the perceived \textit{lounge comfort}, available \textit{catering} quality, nice \textit{staff service} and the area \textit{cleanliness}. A probably expected observation is that the perceived \textit{catering} quality is highly influenced by the availability of \textit{beverages}. An interesting finding in lounge reviews is that the sentiment not only correlates with the overall traveler satisfaction but also with the various rating features that denote specific services provided in lounges.

\vspace{2mm}
\noindent \textbf{Airline reviews.}
With respect to airline reviews, the top influencing rating feature is \textit{value-for-money}. We also find that how a traveler perceives the \textit{cabin staff} may be influenced by the \textit{seat comfort} and the availability of \textit{food and beverages}. 
The extracted sentiment from the review text mostly correlates with the \textit{overall rating}, being here the second best correlating feature and as such a strong signal for traveler satisfaction.

\vspace{2mm} \noindent \textbf{Seat reviews.}
With respect to the overall satisfaction of a traveler's seat, the best correlating features are the \textit{legroom}, \textit{width}, \textit{recline} and \textit{aisle space}. Looking at how this distinctive features correlate with each other also suggests, although somehow intuitive, that a traveler's available personal space is the most important factor when sitting on a plane. Another interesting observation is that how a traveler is satisfied with the available \textit{seat storage} is highly influenced by the availability of a \textit{power supply}. The review sentiment, similar as in the case of lounge and airline reviews, is again a strong indicator for the traveler satisfaction denoted by the overall rating.  

\vspace{2mm} \noindent \textbf{Extracting review topics.}
With respect to clustering and cluster labelling, in Figure \ref{fig:topics}, we report a snapshot of our preliminary results using the Suffix Tree Clustering (STC) approach. 
By utilizing STC, additional textual features (i.e., cluster labels) can be extracted from the review content. For example, we see in Figure \ref{fig:topics} that travelers write about \textit{boarding} time when experiencing negative traveler satisfaction, which in turn results into a negative review about the specific airline. %
On the contrary, travelers seem to be satisfied with airports when, for example, a smooth \textit{immigration} is ensured and when \textit{gates} are labeled well and easy to reach. Consequently, existing rating schemes could be extended with such cluster labels if they reflect recurring points of discussion in textual reviews.

\section{Predicting \\ Traveler Satisfaction}
\label{sec:rp}
In this section, we aim to address our second research question (\textit{RQ2}) in order to determine the features that can be exploited to predict the final traveler satisfaction. Therefore, we formulate the prediction task as a binary classification problem. Given that reviews are marked as either \textit{positive} or \textit{negative} traveler satisfaction by means of the \textit{Would you recommend this airline/airport?} checkbox of Skytrax, we aim to predict this outcome using the available rating and textual features.

\subsection{Methodology} 
We performed our experiments using several standard classification algorithms (e.g., NaiveBayes, C4.5, Random Forest, CART, etc. \cite{breiman1984classification,zhao2008comparison}) provided by the
popular machine learning tool WEKA \cite{hall2009weka}. 
In this work, however, we report the results of only one distinguished algorithm, namely Hoeffding Tree.

\begin{figure}[t!]
	\includegraphics[width=0.40\textwidth]{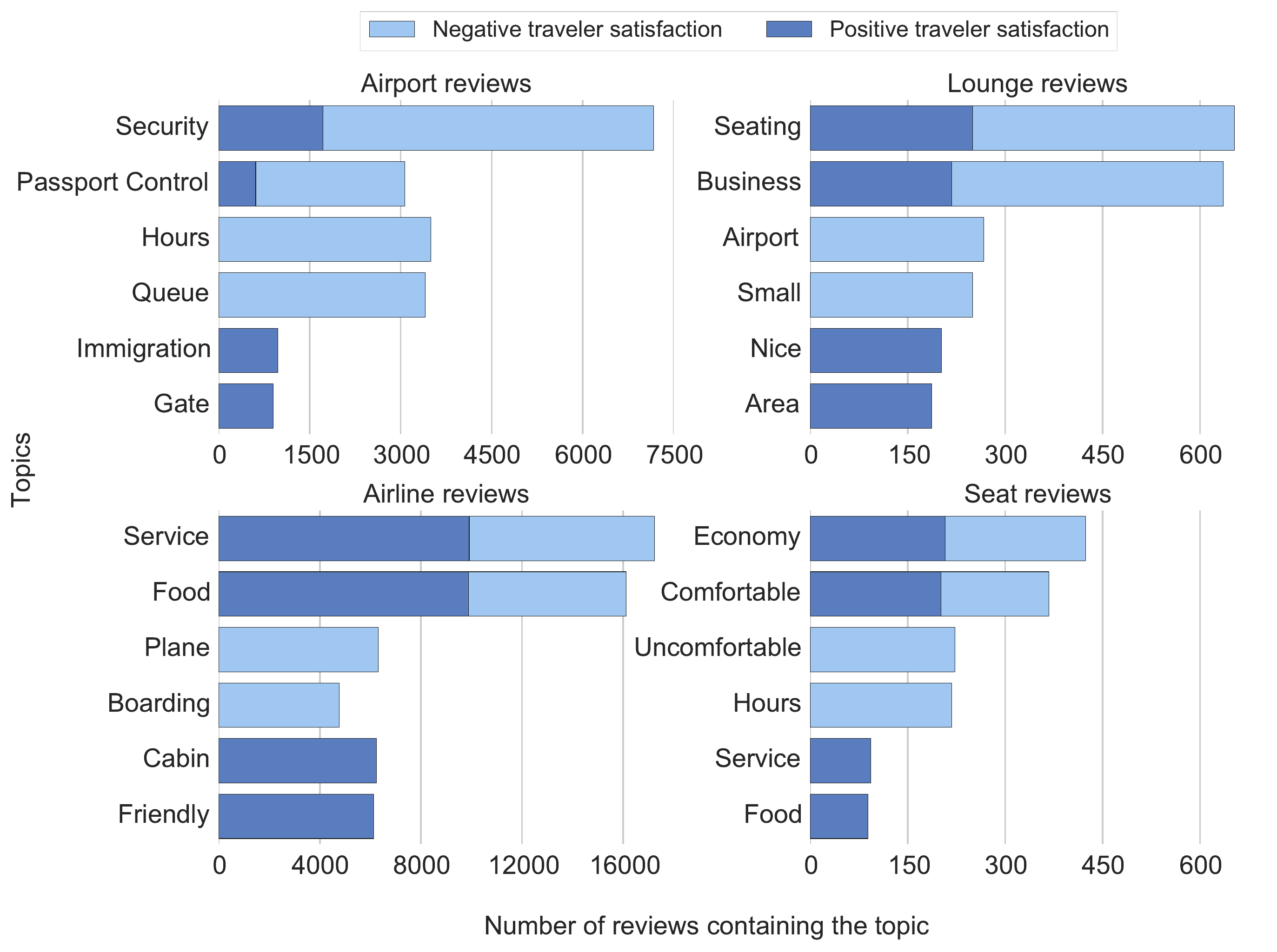}%
     \caption{Snapshot of our preliminary clustering and cluster labeling analysis of the review content using the Suffix Tree Clustering (STC) approach. The extracted topics are grouped by having a positive or negative traveler satisfaction denoted by the \textit{Would you recommend this airline/airport?} checkbox of Skytrax.\vspace{-4mm}}
    \label{fig:topics}
\end{figure}

Introduced by Domigos and Hulten \cite{domingos2000mining}, the Hoeffding Tree algorithm is an incremental decision tree learner for large data streams. The tree itself tracks only attribute statistics in its leafs and uses it to grow and make classification decisions for incoming data. When sufficient statistics have accumulated at each leaf, a node-splitting approach determines whether a node-split should happen and the leaf should be replaced with a new decision node.
We chose this algorithm due to its practical advantage for real-time data mining \cite{hulten2001mining}.

In order to evaluate the classification performance, we sorted the reviews of the four categories in chronological order and used
the 20\% most recent reviews for testing and the rest for training. Next, using each of the four training sets, we examined whether the final user satisfaction of a target review from the corresponding test set could be predicted. 
With this procedure, we aim to simulate a real-world environment in which future reviewing behavior should be predicted based on past reviews. To determine the best performing features for traveler satisfaction prediction, we trained and evaluated the classification model in the following three settings.

Firstly, for each single rating feature, we created a separate classifier and evaluated its performance. Secondly, we combined the best performing rating features identified in our correlation analysis of \textit{RQ1} to create a classification model. Specifically, we incrementally removed the lowest correlating feature until we found the best performing combination. Thus, we report the combination of features with a correlation value higher than 0.3 (i.e., overall rating, queuing, airport shopping and terminal cleanliness in the case of airport reviews as seen in Figure \ref{fig:correlation}). Thirdly, we trained a model solely based on the inferred review text sentiment.
In order to finally quantify the prediction performance, we used a set of well-known information retrieval metrics. In particular, we report the prediction accuracy by means of the F1-score (F1) and Area Under the ROC curve (AUC) \cite{powers2011evaluation}.

\begin{table}[t!]
\centering
\scalebox{0.88}{
\begin{tabularx}{0.48\textwidth}{X||C{1.5cm}|C{1.5cm}}
\hline \hline
\multicolumn{3}{c}{\textbf{Airport reviews}} \\ \hline\hline
Feature                 &   F1 & AUC   \\ \hline
Overall                 &     \textbf{0.963}   & 0.948      \\ 
Queuing                 &     0.869    & 0.875     \\ 
Airport shopping        &     0.859    & 0.876      \\ 
Terminal cleanliness    &     0.828    & 0.814      \\ 
Terminal seating        &     0.791    & 0.534      \\ 
Food beverages          &     0.792    & 0.530      \\ 
WiFi connectivity       &     0.774    & 0.519   \\ 
Terminal signs          &     0.800    & 0.502      \\ 
Airport staff           &     0.678    & 0.499      \\ \hline
Combination             &     0.962     & \textbf{0.973}     \\ \hline
Airport Sentiment       &     0.719    &  0.715       \\ 
\hline \hline
\multicolumn{3}{c}{\textbf{Lounge reviews}} \\ \hline \hline
Feature                 &   F1 & AUC           \\ \hline
Overall                 &   0.834    & 0.878    \\ 
Comfort                 &   0.762    & 0.839     \\ 
Staff service           &   0.768    & 0.819   \\ 
Bar beverages           &   0.783    & 0.838    \\ 
Catering                &   0.783    & 0.829    \\ 
Cleanliness             &   0.773    & 0.817    \\ 
Washrooms               &   0.750    & 0.826     \\ 
WiFi                    &   0.743    & 0.795    \\ 
\hline
Combination             &   \textbf{0.837}    & \textbf{0.884}    \\ \hline
Lounge Sentiment        &   0.773     &  0.822   \\ 
 \hline \hline
\multicolumn{3}{c}{\textbf{Airline reviews}} \\ \hline \hline
Feature                 &   F1 & AUC          \\ \hline
Overall                 & 0.838     & 0.971     \\ 
Value money             & \textbf{0.863}     & 0.940     \\ 
Cabin staff             & 0.794     & 0.884     \\ 
Seat comfort            & 0.750     & 0.843    \\ 
Food beverages          & 0.741     & 0.827     \\ 
Inflight entertainment  & 0.693     & 0.754    \\ 
Ground service          & 0.622     & 0.533   \\ 
WiFi connectivity       & 0.615     & 0.509   \\ 
\hline
Combination             &  0.844    & \textbf{0.974}    \\ \hline
Airlne Sentiment        & 0.839     &   0.896      \\ 
 \hline \hline
\multicolumn{3}{c}{\textbf{Seat reviews}} \\ \hline \hline
Feature  &   F1 & AUC          \\ \hline
Overall      &  \textbf{0.939}     & \textbf{0.985}     \\ 
Seat legroom & 0.872     & 0.919     \\ 
Seat width   & 0.847     & 0.890   \\ 
Aisle space  & 0.840     & 0.895    \\ 
Seat recline &  0.802     & 0.855     \\ 
Viewing TV   &  0.730     & 0.759      \\ 
Seat storage & 0.711     & 0.576     \\ 
Power supply &  0.647     & 0.529     \\ 
\hline 
Combination  & 0.917     & 0.981    \\ \hline
Seat Sentiment              &  0.812    &   0.849      \\ 
 \hline \hline
\end{tabularx}
}
\caption{Classification results using the Hoeffding Tree algorithm for each of the four review categories. The accuracy performance of each single rating feature is reported, as well as the performance when the rating features are combined. Additionally, we report the accuracy, which is achieved by only using review text sentiment as the sole feature. All results are reported by means of the F1-score and AUC (\textit{RQ2}).\vspace{-4mm}}
\label{tab:info}
\end{table}

\subsection{Results}
In this section, we present our prediction results of the Hoeffding Tree algorithm for the individual rating features, the combination of rating features, the review text sentiment as well as a discussion on runtime considerations.

\vspace{2mm} \noindent \textbf{Individual rating features.}
Our prediction results based on the review categories are shown in Table \ref{tab:info}. In general, we find strong accuracy performance in predicting the traveler satisfaction using the overall rating feature (e.g., F1 = 0.963 for airport reviews). Furthermore, the performance of rating features that have shown a high correlation with the overall rating (see \textit{RQ1}) also perform reasonably well in terms of satisfaction prediction. For example, using the \textit{value-for-money} feature (F1 = 0.863) in airline reviews provides higher prediction accuracy than using the overall rating (F1 = 0.838).

This finding indicates that travelers perceive the received value for the spent money as the strongest influence on their final satisfaction with a flight. In contrast, we observe that features with a weak correlation to the overall rating also reach low AUC estimates below 0.6, which is only slightly above random guessing.

\vspace{2mm} \noindent \textbf{Combination of rating features.} Overall, the combination of rating features results in strong prediction results with respect to F1-score and AUC. The best performance is achieved with lounge reviews. While being the second best performing feature in airport, airline and seat reviews, the prediction accuracy is still high and does not differ that much from the best performing feature. In case of airport and airline reviews the feature combination even shows the best AUC performance.

\vspace{2mm} \noindent \textbf{Review text sentiment.} Compared to other rating features, review text sentiment is a competitive feature when predicting traveler satisfaction. For example, we can observe that for airline reviews, the sentiment is the third best performing feature (F1 = 0.839), outperforming even the overall rating. 

\vspace{2mm} \noindent \textbf{Runtime considerations.}
When training and testing the different classification approaches, we experienced the best accuracy performance for the Hoeffding Tree algorithm. Moreover, we found a maximum model training runtime of $0.06$ seconds for this classifier in case of the rating feature combination for airline reviews. %
This clearly underpins our choice for the Hoeffding Tree classifier since runtime is crucial when new review data, which is mined from online portal, should be instantly included in the classification process. As these results show, Hoeffding Tree is able to build a competitive model in a reasonable time and enables incremental data updates with no need for re-training the complete model, which is crucial for real-time data mining applications \cite{hulten2001mining}.

\vspace{-2mm}
\section{Conclusion and Future Work} 
\label{sec:con}
In this paper, we discussed how online reviews can be an important source of information to explain (\textit{RQ1}) and predict (\textit{RQ2}) traveler satisfaction. Therefore, we used data crawled from the Skytrax portal in order to show that rating features such as airport \textit{queuing} time, lounge \textit{comfort}, airline \textit{cabin staff} quality and seat \textit{legroom} size highly contribute to the overall traveler satisfaction. Moreover, we found a strong correlation between review text sentiment and the final traveler satisfaction (\textit{RQ1}). Based on these findings, we trained several classifiers and reported the results of the Hoeffding Tree algorithm, which not only provides strong accuracy performance but also provides practical advantage when mining data in real-time. Summarized, we found not only that traveler satisfaction can be indeed predicted with high accuracy but also that textual features such as the extracted sentiment bear great potential in explaining and predicting traveler satisfaction (\textit{RQ2}). 

We believe that having strong accuracy and runtime performance is especially beneficial for practical purposes where it is the aim to continuously mine and predict traveler satisfaction using online reviews. As such, our proposed methods and findings of this work should be of interest for researchers in the area of modeling and predicting user satisfaction based on review data on the Web.

\vspace{2mm} \noindent \textbf{Limitations and future work.} In our opinion, a limitation of this work is the lack of a direct comparison with other incremental classifiers such as Incremental Tree Induction (i.e., ITI, the successor of ID5R) or FlexDT (Flexible Decision Tree based on fuzzy logic). As such, we plan for future work to conduct an extensive comparison between different incremental classifiers when mining and predicting user satisfaction using online reviews. Moreover, we want to continue our preliminary investigations of extracting review topics presented in Figure \ref{fig:topics} by further analyzing the textual content of online airline reviews. In this respect, we plan to extend the topic extraction process conducted on the review text with additional approaches like TextRank (one of the most well-known graph-based approaches for keyphrase extraction) and Topical PageRank (runs TextRank  multiple times for topics induced by a Latent Dirichlet Allocation from the text). Therefore, it is not only our aim to uncover additional features that help in explaining traveler satisfaction but also to integrate them in the process of predicting traveler satisfaction. With respect to our prediction study, we plan to incorporate further approaches known from research on recommender systems such as Collaborative Filtering or Matrix Factorization.

\vspace{2mm} \noindent \textbf{Acknowledgments.}
The authors would like to thank Dieter Theiler and Simone Kopeinik for valuable comments on this work. This work is supported by the Know-Center and the EU-funded project Learning Layers under grant agreement 318209. %

\balance
\bibliographystyle{abbrv}

\begin{thebibliography}{10}

\bibitem{breiman1984classification}
L.~Breiman, J.~H. Friedman, R.~A. Olshen, and C.~J. Stone.
\newblock {\em Classification and Regression Trees}.
\newblock Wadsworth International Group '84.

\bibitem{chatterjee2001online}
P.~Chatterjee.
\newblock Online reviews: do consumers use them?
\newblock {\em Advances in Consumer Research '01}.

\bibitem{chen2004impact}
P.-Y. Chen, S.-y. Wu, and J.~Yoon.
\newblock The impact of online recommendations and consumer feedback on sales.
\newblock In {\em Proc. of ICIS '04}.

\bibitem{chen2003marketing}
Y.~Chen, S.~Fay, and Q.~Wang.
\newblock Marketing implications of online consumer product reviews.
\newblock {\em Business Week '03}.

\bibitem{domingos2000mining}
P.~Domingos and G.~Hulten.
\newblock Mining high-speed data streams.
\newblock In {\em Proc. of ACM SIGKDD '00}.

\bibitem{hall2009weka}
M.~Hall, E.~Frank, G.~Holmes, B.~Pfahringer, P.~Reutemann, and I.~Witten.
\newblock The weka data mining software: an update.
\newblock {\em ACM SIGKDD Explorations Newsletter '09}.

\bibitem{heskett1994putting}
J.~L. Heskett, L.~Schlesinger, et~al.
\newblock Putting the service-profit chain to work.
\newblock {\em Harvard business review '94}.

\bibitem{hulten2001mining}
G.~Hulten, L.~Spencer, and P.~Domingos.
\newblock Mining time-changing data streams.
\newblock In {\em Proc. of ACM SIGKDD '01}.

\bibitem{lee1988thirteen}
J.~Lee~Rodgers and W.~A. Nicewander.
\newblock Thirteen ways to look at the correlation coefficient.
\newblock {\em The American Statistician '98}.

\bibitem{liou2010dominance}
J.~J. Liou and G.-H. Tzeng.
\newblock A dominance-based rough set approach to customer behavior in the
  airline market.
\newblock {\em Information Sciences '10}.

\bibitem{pang2008opinion}
B.~Pang and L.~Lee.
\newblock Opinion mining and sentiment analysis.
\newblock {\em Foundations and trends in information retrieval '08}.

\bibitem{powers2011evaluation}
D.~M.~W. Powers.
\newblock {Evaluation: from precision, recall and F-measure to ROC,
  informedness, markedness and correlation}.
\newblock {\em International Journal of Machine Learning Technology '11}.

\bibitem{saha2009service}
G.~C. Saha and Theingi.
\newblock Service quality, satisfaction, and behavioural intentions: A study of
  low-cost airline carriers in thailand.
\newblock {\em Managing Service Quality: An International Journal '09}.

\bibitem{sambasivam2006advanced}
S.~Sambasivam and N.~Theodosopoulos.
\newblock Advanced data clustering methods of mining web documents.
\newblock {\em Issues in Informing Science and Information Technology '06}.

\bibitem{serrano2015sentiment}
J.~Serrano-Guerrero, J.~A. Olivas, F.~P. Romero, and E.~Herrera-Viedma.
\newblock Sentiment analysis: A review and comparative analysis of web
  services.
\newblock {\em Information Sciences '15}.

\bibitem{suki2014passenger}
N.~M. Suki.
\newblock Passenger satisfaction with airline service quality in malaysia: A
  structural equation modeling approach.
\newblock {\em Research in Transportation Business \& Management '14}.

\bibitem{tiernan2008airline}
S.~Tiernan, D.~L. Rhoades, and B.~Waguespack~Jr.
\newblock Airline service quality: Exploratory analysis of consumer perceptions
  and operational performance in the usa and eu.
\newblock {\em Managing Service Quality: An International Journal '08}.

\bibitem{Vlachos2014}
I.~Vlachos and Z.~Lin.
\newblock {Drivers of airline loyalty: Evidence from the business travelers in
  China}.
\newblock {\em Transportation Research Part E: Logistics and Transportation
  Review '14}.

\bibitem{Yakut2015}
I.~Yakut, T.~Turkoglu, and F.~Yakut.
\newblock Understanding customer's evaluations through mining airline reviews.
\newblock {\em International Journal of Data Mining {\&} Knowledge Management
  Process '15}.

\bibitem{yao2015exploring}
B.~Yao, H.~Yuan, Y.~Qian, and L.~Li.
\newblock On exploring airline service features from massive online review.
\newblock In {\em Proc. of ICSSSM '15}.

\bibitem{Zamir1998}
O.~Zamir and O.~Etzioni.
\newblock Web document clustering: A feasibility demonstration.
\newblock In {\em Proc. of ACM SIGIR '98}.

\bibitem{zhao2008comparison}
Y.~Zhao and Y.~Zhang.
\newblock Comparison of decision tree methods for finding active objects.
\newblock {\em Advances in Space Research '08}.

\end{thebibliography}

\end{document}